\documentclass[useAMS,usenatbib]{mn2e}
\usepackage{pdflscape}
\usepackage{graphicx}
 \usepackage{amsmath} % not originally on MNRAS style, equations shifted...
\usepackage{amssymb}
\usepackage{mathrsfs}

\usepackage{graphics}
\usepackage{graphicx}
\usepackage{wrapfig,epsfig,epsf}
\usepackage{bm}    %%%% boldface in math mode

\usepackage{hyperref}
\title[Logarithmic potential]{Logarithmic potential for the gravitational field\\ of Schwarzschild black holes}

\author[N. I. Shakura and G. V. Lipunova]{N. I. Shakura$^{1,2}$ and G. V. Lipunova$^1$\\
$^1$ Moscow  Lomonosov State University, Sternberg Astronomical Inst., 
Universitetski pr. 13, Moscow 119234, Russia;\\  
$^2$  Kazan Federal University, 18 Kremlevskaya, Kazan, 420008, Russia, {nikolai.shakura@gmail.com}}
\begin{document}
\date{Accepted 2018 August 3. Received 2018 August 1; in original form 2018 May 28}

\pagerange{\pageref{firstpage}--\pageref{lastpage}} %\pubyear{2002}

\maketitle

\label{firstpage}

\begin{abstract}
Approximate gravitational potentials are often used to  describe analytically  the motion of particles near black holes (BHs), as well as to study the structure of an accretion disk. Such 'pseudo-Newtonian' potentials are used with the flat-metric equations.   Here we consider the motion of a free  particle near a non-rotating BH in the context  of an exact `logarithmic' gravitational potential. We  show how the logarithmic potential gives an exact solution  for a  mechanical problem { and present the relativistic Bernoulli equation for the fluid in the Schwarzschild metric.}   
\end{abstract}


\newcommand{\ar}{a_\mathrm{r}}
\newcommand{\boldnabla}{\mbox{\boldmath$\nabla$}}
\newcommand{\br}{b_\mathrm{r}}
\newcommand{\Bt}{B_{\small\mbox{\textsc{t}}}}
\newcommand{\D}{\mathrm{d}}
\newcommand{\de}{\partial}
\newcommand{\EBV}{E(\mbox{B}-\mbox{V})}
\newcommand{\Mus}{\mbox{GS\,1124--683\protect}}
\newcommand{\ho}{h_0}
\newcommand{\kapf}{\varkappa_\mathrm{ff}}
\newcommand{\kapt}{\varkappa_{\small\mbox{\textsc{t}}}}
\newcommand{\kappaT}{\kappa_{\small\mbox{\textsc{t}}}}
\newcommand{\Led}{L_{\mathrm{Edd}}}
\newcommand{\ls}{\left(}
\def\Mon{{\hbox{A\,0620--00}}}{}%
\newcommand{\mx}{m_\mathrm{x}}
\newcommand{\NH}{N(\mathrm{HI})}
\newcommand{\nutc}{\nu_\mathrm{t}^\mathrm{c}}
\newcommand{\nut}{\nu_\mathrm{t}}
\newcommand{\omegak}{\omega_{\small\mbox{\textsc{k}}}}
\newcommand{\planck}{{\small\mbox{\textsc{p}}}}
\newcommand{\pc}{P_\mathrm{c}}
\newcommand{\ps}{\right)}
\newcommand{\Qo}{Q_0}
\newcommand{\rg}{r_{\mathrm{g}}}
\newcommand{\Rg}{R_{\mathrm{g}}}
\newcommand{\Rs}{R_{\mathrm{g}}}
\newcommand{\rhoc}{\rho_\mathrm{c}}
\newcommand{\rin}{r_{\mathrm{in}}}
\newcommand{\rout}{r_\mathrm{out}}
\newcommand{\rISCO}{r_\mathrm{ISCO}}
\newcommand{\sigmasb}{\sigma_{\small\mbox{\textsc{sb}}}}
\newcommand{\sigmat}{\sigma_{\small\mbox{\textsc{t}}}}
\newcommand{\sigmo}{\Sigma_0}
\newcommand{\slfrac}[2]{\left.#1\middle/#2\right.}
\newcommand{\shsc}{\cite{shakura-sunyaev1973}}
\newcommand{\tauf}{\tau_\mathrm{ff}}
\newcommand{\taut}{\tau_{\small\mbox{\textsc{t}}}}
\newcommand{\tc}{T_\mathrm{c}}
\newcommand{\tobs}{\Delta t_\mathrm{ob}}
\newcommand{\teff}{T_\mathrm{eff}}
\newcommand{\Tsymbol}{{\small\mbox{\textsc{t}}}}
\newcommand{\Zo}{Z_0}
\def\wrf{W_{r\varphi}}
\def\aap{A\&A{ }}%
\def\apjl{ApJ{ }}%
\def\apj{ApJ{ }}%
\def\aj{AJ{ }}%
\def\apjs{ApJS{ }}%
\def\mnras{MNRAS{ }}%
\def\apss{Ap\&SS{ }}%
\def\araa{ARA\&A{ }}%
\def\prd{Phys.~Rev.~D{ }}%
\def\nat{Nature{ }}%
\def\pasp{PASP{ }}%
\def\azh{АЖ{ }}%
\def\pasj{PASJ{ }}%

\section{Introduction}

{ In the Newton celestial mechanics, a gravitational potential is one of the basic concepts. In the General Relativity (GR) there is generally no such concept as a gravitational potential. In some special cases, however, it is possible to use such a concept, as we show in this work. This gravitational potential is different from what is usually termed as a pseudo-Newtonian potential. }

To  describe analytically and in a simple way the dynamics of particles near a BH,
as well as to study the structure of an accretion disk, approximate 
approaches are frequently used. For example, it is common to utilize  pseudo-Newtonian gravitational potentials in the equations written in the flat ,etric. For a non-rotation BH, the potential by \citet{pacz-wiita1980} is used (hereafter, `PW potential'). For a rotating black hole, \cite{artemova+1996} proposed a formula for a pseudo-Newtonian gravitational force acting on particles near Kerr BH.

Here we consider  a non-rotating BH  and a `logarithmic' gravitational potential. This gravitational potential, together with an allowance for the curvature of the space-time, provide the laws of motion for a free particle, which are identical to those  derived in the General Relativity (GR).

In Sect.~\ref{s.potentials} the pseudo-Newtonian gravitational potentials are very briefly reviewed. { We introduce the logarithmic potential in Sect.~\ref{s.log_potential}.}
In Sect.~\ref{s.solution} we consider the equation of motion  
of a particle in a curved space-time and derive the conserved value of  energy. We obtain the law of  motion for the logarithmic potential and consider its consequences in Sect.~\ref{s.velocities}. { The relativistic Bernoulli equation for a stationary fluid around a Schwarschild BH is derived in Sect.~\ref{s.bernoulli}. }

\section{Pseudo-Newtonian Gravitational potentials}
\label{s.potentials}

Near a black hole (BH), the curvature of the space-time is a decisive factor affecting the structure 
of an accretion disc. 
For a non-rotating black hole, the radius of the innermost stable circular orbit $r_\mathrm{ISCO}=3\, \Rs$, where the Schwarzschild radius $\Rs$ is the event horizon of a non-rotating black hole.
$$
\Rs = 2\,G\,M/c^2\, .  
$$

To approximate effects of the GR in the vicinity of  a non-rotating black hole, the Paczynski--Wiita potential can be used \citep{pacz-wiita1980}:
\begin{equation}
\Phi_\mathrm{PW} = - \frac {G\, M} {r-\Rs}\, .
\label{eq.Phi_PW}
\end{equation}

For free particles in circular orbits, the velocities can be found from the radial component of the Navier-Stokes equation
\begin{equation}
	\frac{v_\varphi^2}{r} = \frac{ \mathrm{d} \Phi}{\mathrm{d}  r}\, .
\label{eq.vphi2_over_r}
\end{equation}
As a result, one obtains  the orbital velocity
$$
\frac{v_\varphi^\mathrm{PW}}{c} = {\frac{1}{\sqrt{2}}}\,  \frac{\sqrt{r\, \Rs}}{(r-\Rs)} \, ,
$$
and the specific angular momentum of a test particle in the Paczynski-Wiita potential:
\begin{equation}
h^\mathrm{PW} = v_\varphi^\mathrm{PW} \, r =  \sqrt{\frac{G\, M\, r}{(1-\frac{\Rs}{r})^2}}\, .
\label{eq.glava1_h_pacz}
\end{equation}

The modified potential \eqref{eq.Phi_PW} is often used in hydrodynamic and magneto-hydrodynamic numerical codes, since it approximates quite well the curvature effects of the space-time metric around a Schwarzschild black hole~(\citet{Yuan-Narayan2014}; e.g., \citet{Ohsuga-Mineshige2011,Jiang+2014}). Other approximate potentials, in particular such applicable to the case of rotating black holes, can be found in~\citet{artemova+1996,kfm_book1998,Witzany+2015}.

\section{Logarithmic potential}
\label{s.log_potential}

To describe the relativistic motion in the vicinity of a Schwarzschild black hole we may use the following `logarithmic' potential~\citep{landau-lifshitz-e-2-1975,
thorne_et1986}:
%% paragraph 88 of Landau & Livshitz
 \begin{equation}
 \Phi =\frac{c^2}{2}\, \ln \left(1-\frac{\Rs}{r}\right) =  {c^2}\, \ln \sqrt{1-\frac{\Rs}{r}}\, .
 \label{eq.Phi_log}
\end{equation}

Below, we will show how the logarithmic potential gives an exact solution  for a  mechanical problem. This will require consideration of the space-time curvature near a Schwarzschild BH.

Note that \citet{artemova+1996} treated the logarithmic potential as a pseudo-Newtonian potential and this provided an approximate result, with an order of accuracy comparable to that of the PW potential.

\section{Equation of motion with logarithmic potential}
\label{s.solution}
\begin{figure}
\centering
\includegraphics[angle=0,width=0.4\textwidth]{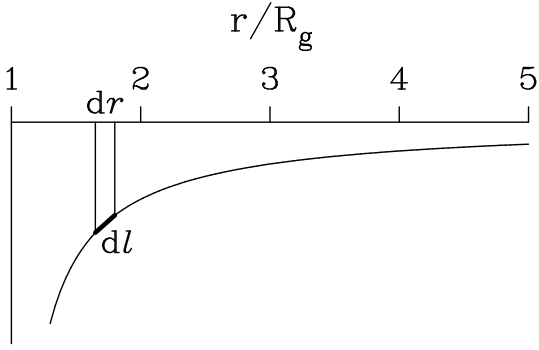}
\caption{
Illustration of the  `shrinking' of a coordinate element $\mathrm{d} r$, corresponding to an element of distance $\mathrm{d} l$, measured by a local static or a fiducial observer (`FIDO' of~\citet{thorne_et1986}).\label{fig.gl1.dl_dr}
}
\end{figure}

Let us write down the Schwarzschild stationary metric as the square of an interval between two events separated in time and space:
$$
 d s^2 = - ( 1-\Rs/r)\, dt^2 +  ( 1-\Rs/r)^{-1}\, dr^2 +
 r^2 (d \theta +\sin^2 \theta \, d\varphi)\, .
$$
Here, $t,~r,~\theta$, and $~\varphi$ are the Schwarzschild coordinates. Due to the curvature of the  space-time near a black hole,
the distance element $\mathrm{d}l$ along the radius, as measured by a local observer, is longer than the corresponding
coordinate element $\mathrm{d}r$ (see Fig.~\ref{fig.gl1.dl_dr}):
$$
\mathrm{d} l = \frac{\mathrm{d} r}{\sqrt{1-\Rs/r}}\, .
$$

Inherited by \eqref{eq.Phi_log},  $\sqrt{1-\Rs/r}$ is a lapse function in the Schwarzschild metric. It determines the redshift of a signal emitted from the vicinity of a black hole and
the difference between  two  time  intervals,  one  of  which,  $\D  t$,  is  measured  at  infinity  and  the  other,  $\D\tau_{l}$,  by  an  observer  in  the  local  stationary  reference  frame:
\begin{equation}
\mathrm{d} \tau_{l}/ \mathrm{d} t = \sqrt{1-\Rs/r}\, .
 \label{eq.tau_loc_tau_inf}
\end{equation}
The time measured in the frame of a moving particle is related to the time measured by a local stationary observer as 
\begin{equation}
\mathrm{d} \tau_\mathrm{p}/  \mathrm{d} \tau_{l}  =  \sqrt{1-{v}^2/c^2}\, .
  \label{eq.tau_loc_tau_p}
\end{equation}

 Let us consider a relativistic particle with the rest mass $m_o$. Its momentum ${\pmb p}$ and energy $E_\mathrm{local}$,  relative to the local stationary observer, are
$$
{\pmb p} = \frac {m_o\,{\pmb v}}{\sqrt{1-{v}^2/c^2}}\,    \quad \mbox{and} \quad
E_\mathrm{local} = \frac{m_o\,c^2}{\sqrt{1-{v}^2/c^2}}~,
$$
respectively, where the square velocity $v^2 = v_r^2 + v_\varphi^2$ for particles moving in the equatorial plane. 

We may also introduce the notion of `energy at infinity' $E$. This value remains unchanged along the particle trajectory. Let us determine it.

Consider a particle travelling past a stationary observer who is located at a distance from a black hole.
The equation of particle motion in the reference system of this observer can be written as follows\citep{landau-lifshitz-e-2-1975}:
\begin{equation}
\frac{\mathrm{d}{ \pmb p}}{\mathrm{d} \tau_{l}} = -
\frac{m_o}{\sqrt{1-{ v}^2/c^2}}\, \pmb \nabla \Phi\, .
\label{eq.ur_dvizheniya}
\end{equation}

As it is done in mechanics, the energy of a particle can be found from the equation of motion by  multiplying scalarly Eq.~\eqref{eq.ur_dvizheniya} by $\pmb{v}$:
$$
{\pmb v} \, \frac{\mathrm{d}}{\mathrm{d} \tau_{l}} \, \left(
\frac{m_o { \pmb v} }{ \sqrt{1-v^2/c^2}} \right) = -
\frac{m_o\, {\pmb v}}{\sqrt{1-{ v}^2/c^2}}\, \pmb \nabla \Phi
$$
or, noting that the potential $\Phi$ is spherically symmetric,
\begin{equation}
{\pmb v} \, \frac{\mathrm{d}}{\mathrm{d} \tau_{l}} \, \left(
\frac{m_o { \pmb v} }{ \sqrt{1-v^2/c^2}} \right) = -
\frac{m_o\, {\pmb v}\, \pmb{e}_{r} } {\sqrt{1-{ v}^2/c^2}}\, \frac{  \mathrm{d} \Phi}{\mathrm{d} l}\, ,
\label{eq.ur_dvizheniya1}
\end{equation}
where $\pmb{e}_{r}$ is a unit radial vector in the Cartesian reference system of the local observer.  
Further, we differentiate the left-hand part of Eq. \eqref{eq.ur_dvizheniya1}:
$$
\frac12\, \frac{m_o}{\sqrt{1-v^2/c^2}} \,\frac{\mathrm{d}{  v^2}}{\mathrm{d} \tau_l} + \frac12\, \frac{m_o\, v^2/c^2 \,}{({1-{ v}^2/c^2})^{3/2}} \, \frac{\mathrm{d}{  v^2}}{\mathrm{d} \tau_l}  = -
\frac{m_o\,{\pmb v}\, \pmb{e}_{r}}{\sqrt{1-{ v}^2/c^2}}\, \frac{  \mathrm{d} \Phi}{\mathrm{d} l}\, .
$$

When multiplying this by $({1-{ v}^2/c^2})^{3/2}$, cancelling out the two equal terms with opposite signs 
in the left-hand part of the equation and using the equality $v_{r} = {\mathrm{d} {l}}/{\mathrm{d} \tau_{l}}$ for the radial velocity,  we obtain
$$
\frac12 \, \frac{\mathrm{d}}{\mathrm{d} \tau_{l}}  \,(1 -v^2/c^2)=
(1 -v^2/c^2)\, \frac{\mathrm{d} l}{\mathrm{d} \tau_{l}} \, \frac{\mathrm{d}}{\mathrm{d} l} \ln (1-\Rs/r)^{1/2} \, ,
$$
which is equivalent to the following equation
$$
 {\frac{\mathrm{d}}{\mathrm{d}\tau_{l}}\,\ln (1-v^2/c^2)  } = \frac{ \mathrm{d}}{\mathrm{d}  \tau_{l}}\, \ln (1-\Rs/r)\, .
$$

As a result, we obtain the following relationship: 
$$
(1-\Rs/r)\Big/(1-{v}^2/c^2  ) = const.
$$
Hence, the value 
\begin{equation}
 E = \frac{m_o\,c^2}{\sqrt{1-{ v}^2/c^2}}\, \sqrt{1-\frac{\Rs}{r}} =   E_\mathrm{local}\, \sqrt{1-\frac{\Rs}{r}} =const\, ,
 \label{eq.gl1_E_infinity}
\end{equation}
does not change for a freely moving particle, while the locally measured energy $E_\mathrm{local}$ varies in the gravitational field of the black hole. This value $E$ is termed `energy-at-infinity'~\citep{thorne_et1986}. In GR, the value $E$ corresponds to the time component of the 4-vector impulse~\citep{landau-lifshitz-e-2-1975}.

For a photon, the rest mass of which is $m_o=0$, Eq. \eqref{eq.gl1_E_infinity} yields a relation between its frequency $\nu_{o}$ in the reference system of the local observer, and its frequency detected at infinity $\nu_\infty = \nu_o  \sqrt{1-\Rs/r}$. This relation  describes the redshift effect.

In the non-relativistic approximation, energy $\mathscr{E}_\mathrm{N}$  of a particle has the well-known form 
\begin{equation}
E-m_o\,c^2 \equiv \mathscr{E}_\mathrm{N} =
m_o\,v^2/2 - m_o\,G\, M/r\, .
\label{eq.energy_cons_newt}
\end{equation}
{ Let us underline a difference between post-Newtonian approximations and the approach that we use here. A pseudo-Newtonian potential enters   \eqref{eq.energy_cons_newt} in place of the Newtonian potential and is a term of a sum, while the exact expression for the conserved energy \eqref{eq.gl1_E_infinity} is a product of two terms.}

\section{Velocities and binding energy}\label{s.velocities}

Let us now determine the components of the particle velocity in the equatorial plane.  A freely moving particle   with mass $m_o$ in the  spherically-symmetrical gravitational potential keeps its angular momentum unchanged~\citep{landau-lifshitz-e-2-1975}
\begin{equation}
h_\mathrm{p} = \frac{m_o\,v_\varphi\, r}{\sqrt{1-{ v}^2/c^2}}\, . 
\label{eq.rel_mom_impulsa}
\end{equation}
 When taking into consideration that $v^2 = v_r^2 + v_\varphi^2$, Eqs. \eqref{eq.gl1_E_infinity} and \eqref{eq.rel_mom_impulsa} yield
\begin{equation}
\frac{v_r^2}{c^2} = 1 - \frac{m_o^2\, c^4}{ E^2} \, \left(\frac{h_\mathrm{p}^2}{ r^2\,m_o^2\,c^2} +1 \right) \, \left( 1- \frac{\Rs}{r} \right).
\end{equation}
Multiplying by a factor $E^2/(m_o^2\, c^4)$ and using equations \eqref{eq.tau_loc_tau_p} and \eqref{eq.gl1_E_infinity} together with the relation
$$
\frac{v_r^2}{c^2}  = \frac{1}{c^2}\,\left(\frac{\mathrm{d} r }{\mathrm{d} \tau_\mathrm{p}} \right)^2 \, \frac{m_o^2\, c^4}{E^2}\, ,
$$
we may rewrite the last expression. As a result, we obtain the law of motion for a particle with energy
$E$, which is identical to the exact solution in GR, see~\citet{sha-teu1983}:
$$
\frac{1}{c^2}\, \left(\frac{\mathrm{d} r }{\mathrm{d} \tau_\mathrm{p}} \right)^2 = \frac{E^2}{m_o^2\, c^4}  -  \left(\frac{h_\mathrm{p}^2}{ r^2\,m_o^2\,c^2} +1 \right) \, \left( 1- \frac{\Rs}{r} \right)  \, .
$$

Note that in the approximation of a Newtonian potential, this law of motion looks like:
$$
v_r^2 = 
\frac {2}{m_o}\, \left(\mathscr{E}_\mathrm{N} + m_o\,\frac{G\,M}{r}\right) - \frac{h_\mathrm{N}^2 } { r^2\,m_o^2}\, ,
$$
where $h_\mathrm{N} = m_o\,v_\varphi\, r = const$.

Let us consider particles moving in circular orbits around a Schwarzschild black hole. For such motion,
both $v_r$ and ${\mathrm{d} r }/{\mathrm{d} \tau_\mathrm{p}}$ become zero. For the sake of convenience, we may introduce an effective potential
$$
V(r) = \left(\frac{h_\mathrm{p}^2}{ r^2\,m_o^2\,c^2} +1 \right) \, \left( 1- \frac{\Rs}{r} \right)\, .
$$
For circular orbits, the first derivative of this potential becomes zero (the potential has an extremum).
The system of equations 
$$
\frac{\mathrm{d} r }{\mathrm{d} \tau_\mathrm{p}} =0\, ,\qquad\mbox{}\qquad \frac{\partial V(r)}{\partial r} = 0 
$$
yields the following angular momentum in a circular orbit:
\begin{equation}
 h_\mathrm{p}^2 = \frac{m_o^2\, r\,\Rs\, c^2}{2-3\Rs/r}\, .
\label{eq.hkvadrat}
 \end{equation}
After squaring~\eqref{eq.rel_mom_impulsa} and using \eqref{eq.hkvadrat},  we obtain  the tangential velocity as measured by the local observer
\begin{equation}
\frac{v_\varphi}{c} = \frac1{\sqrt{2}} \sqrt{ \frac{\Rs}{r - \Rs} }\, .
\label{eq.gl1_v_phi_log_Phi}
\end{equation}

For the local observer, the angular velocity of a particle is
\begin{equation}
 \omega_{l} =  \frac{v_\varphi}{r} = \frac{c}{\sqrt{2}\,r} \sqrt{ \frac{\Rs}{r - \Rs} }\, .
\end{equation}
Using time-dilation \eqref{eq.tau_loc_tau_inf}, we obtain the angular velocity measured by an observer at infinity:
\begin{equation}
  \omega =  \frac{c\,\sqrt{\Rs}}{\sqrt{2}\,r^{3/2}} = \frac{\sqrt{G\, M}}{r^{3/2}}\, ,
\end{equation}
that is, the classical expression following from Kepler's law.

According to the Rayleigh criterion~\citep{Rayleigh1917}, stable circular orbits cannot exist where $ \mathrm{d} h_\mathrm{p}/\mathrm{d} \mathrm{r} < 0$.  This criterion implies that the innermost stable circular orbit has a radius $r_\mathrm{ISCO} = 3\, \Rs$.

When substituting the velocity $v_\varphi= c/2$, which corresponds to $r_\mathrm{ISCO}$, into \eqref{eq.gl1_E_infinity}, we determine the energy of a particle rotating in the last possible stable orbit. The energy of this particle, $E=m_o\, c^2 \,2\sqrt{2}/3$, is less than its rest energy at infinity, $m_0\, c^2$. This means that when a particle moves from infinity towards the Schwarzschild black hole, i.e. in the process of accretion, the released energy is $(m_0\, c^2-E)\approx 0.0572 \,m_0\, c^2$. Thus, the energy conversion efficiency in the accretion process onto a non-rotating black hole is equal to $\sim$ 6\%. A calculation using the Kerr metric shows that the binding energy of the particles is the largest for a maximally rotating black hole and equals to $1-\sqrt{1/3}\approx0.423$ times the rest energy~\citep{kfm_book2008}.

Extracting the square root of \eqref{eq.hkvadrat}, we find the specific angular momentum of a particle in circular orbit in the Schwarzschild metric:
\begin{equation}
 h = \frac{h_\mathrm{p}}{m_o} =
 \frac{\sqrt{G\,M\,r}}{\sqrt{1-\frac{3\,G\,M}{c^2\,r}}} \, .
 \label{eq.glava1_h_log}
\end{equation}
\begin{table}
\caption{The normalized binding energy of a particle at the innermost stable circular orbit in different gravitational potentials}
\begin{center}
\begin{tabular}{lr}
 & $(m_0\, c^2-E) / (m_0\, c^2)$ \\
 \hline
 \mbox{Newtonian potential~~}&   1/12 = 0.08(3) \\
 \mbox{Paczynski--Wiita potential~~}&   1/16 = 0.0625 \\
\mbox{ Logarithmic potential} & \\
    \mbox{ ~~~~~~~as a pseudo-Newtonian potential} & 0.096\\
 \mbox{Logarithmic potential} &\\
 \mbox{~~~~~~~in Schwarzschild metric}& $1-2\sqrt{2}/3 \approx 0.0572$ \\
  \hline
\end{tabular}\\
\end{center}
\label{tab_eff}
\end{table}
\begin{figure}
\centering
\includegraphics[angle=0,width=0.45\textwidth]{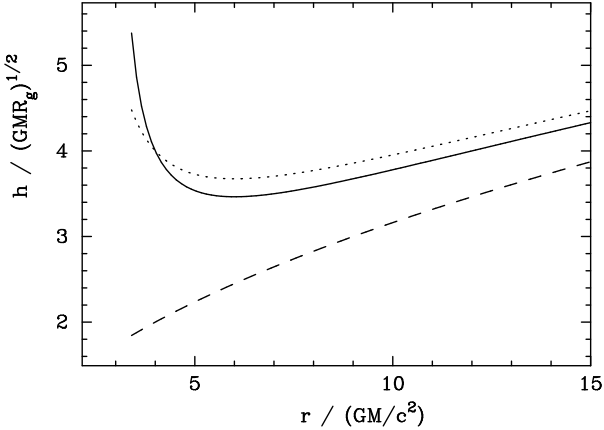}
\caption{Specific angular momentum $h$ of a test particle in the gravitational field of a black hole. The inner radius of the disc is $r_\mathrm{in} = 3 R_\mathrm{g} = 6\, G\,M/c^2$. Solid lines show the dependence in the exact logarithmic potential \eqref{eq.Phi_log}, dotted lines show the same in the Paczynski--Wiita potential, dashed lines -- in the Newtonian approximation.
\label{fig.h_F}
}
\end{figure}

Figure~\ref{fig.h_F} shows the dependence of the specific angular momentum of a test particle on the radius of the orbit in the gravitational field of the black hole.
In addition, the respective dependencies are shown in the Newtonian potential (dashed line) and in the Paczynski--Wiita potential (dotted line).
In the gravitational field of the Schwarzschild black hole, the specific angular momentum $h$ becomes minimum at the radius of the innermost stable circular orbit $6\, G\,M/c^2$. In contrast to the case of the Newtonian potential,  the first derivative of the specific angular momentum,
$\mathrm{d}h/\mathrm{d} r$, vanishes at this radius (see  Fig.~\ref{fig.h_F}).

{ We notice that the innermost stable orbit for the logarithmic potential {\em treated as a pseudo-Newtonian potential} within the classical approach~\citep{artemova+1996} has radius $2\, \Rg$, and the normalized binding energy at this orbit is 0.096. This is an evidently much worse result, comparing to the accuracy provided by the Paczynski--Wiita potential. For the Paczynski--Wiita potential, the radius of the last stable orbit coincides with the GR result, $3\, \Rg$, although the binding energy exceeds by $\sim 9\%$ the exact value (see Table~\ref{tab_eff}).}
    
%The innermost stable orbit is located at $3\, \Rs$ in  both the approximate Paczynski--Wiita potential \eqref{eq.Phi_PW} and the exact potential \eqref{eq.Phi_log}.
%The binding energy in the Paczynski--Wiita potential, however, differs from the exact value (see Table~\ref{tab_eff}).

%Circular orbits exist only down to the radius where $v_\varphi = c$. In the logarithmic potential, the last circular orbit lies at $3\,\Rs/2$, which coincides with the exact value predicted by GR. In the Paczynski--Wiita potential, the last circular orbit is located at $2\, \Rs$.

\section{Relativistic Bernoulli equation}
\label{s.bernoulli}

We have considered above the mechanical characteristics of moving particles. Hydrodynamic equations can be also written for the case of fluid motion in the gravitational filed of a Schwarzschild black hole, using the concept of gravitational potential. Here we consider the Bernoulli equation\footnote{This section was added after we had received essential comments from the anonymous referee.}.  

For an isentropic stationary motion of a fluid we can write an Euler equation in a relativistic form
\begin{equation}
 \gamma\, ({\pmb v}\,  \pmb \nabla) (\gamma\, w\, \pmb v) + c^2\, \pmb \nabla w = - \gamma\, w\, \pmb\nabla \Phi\, .
 \label{eq.isentropic1}
\end{equation}
Here $\omega$ is a `specific' dimensionless enthalpy (per one particle). For $v\ll c$, we have $w = 1+ \frac{w_\mathrm{NR}}{c^2} $, where $w_\mathrm{NR}$ is a non-relativistic enthalpy. For the ideal gas, 
\begin{displaymath} 
w_\mathrm{NR}=  \frac{n}{n+1}  \, \frac{P}{\rho}\,      ,                                                                                                                                                                                                    
\end{displaymath}
where $P$ is the pressure, $\rho$ is the density, $n$ is the adiabatic index ($P \propto \rho^ n$).
Eq.~\eqref{eq.isentropic1} is obtained from a 
a relativistic equation for the energy conservation in a fluid \citep[see~\S~134, Chap.~XV of ][]{landau-lifshitz-hydro-2ed} by adding the term $(- \gamma\, w\, \pmb\nabla \Phi)$ to its right-hand side, which allows for the action of the gravitational force. In this section, we use the following designation: $\gamma = (1-v^2/c^2)^{-1/2}$.

Following the usual rules for transformations with the operator $\pmb\nabla$ \citep{korn1961mathematical}, we re-write the first term in \eqref{eq.isentropic1} as 
\begin{equation}
\gamma \, \pmb v\, (\gamma \pmb v \cdot \pmb\nabla w) + w (\gamma\, \pmb v \pmb \nabla) \, \gamma \pmb v\, .
\label{eq.term1_1}
\end{equation}
Using the rules for a double vector product, the first term in \eqref{eq.term1_1} can be transformed into
$$
\gamma \, \pmb v\, (\gamma \pmb v \cdot \pmb\nabla w)  = \gamma \, \pmb v \times [\gamma \pmb v \times \pmb\nabla w] + \gamma^2\, c^2 \, \pmb\nabla w\,
$$
The second term  of \eqref{eq.term1_1}  can be transformed using another formula of the vector analysis~\citep{korn1961mathematical}:
\begin{equation}
(\gamma\, \pmb v \pmb\nabla) \, \gamma \pmb v = \frac 12\, \pmb\nabla \, \gamma^2\,  v^2 - \gamma\, \pmb v \times [\pmb \nabla \times \gamma \pmb v]
\label{eq.term1_2}
\end{equation}
Now let us convert the first  term in the right-hand side of\eqref{eq.term1_2}:
$$
\frac 12\, \pmb\nabla \, \gamma^2\,  v^2 = \frac 12\, \pmb\nabla\, (c^2 + \gamma^2\, v^2) = \frac 12\, \pmb \nabla \Big(c^2 + \frac{v^2}{1-v^2/c^2}\Big) = 
$$
$$=\frac 12\, \pmb \nabla \frac{c^2}{1-v^2/c^2} = \frac {c^2}{2} \, \pmb\nabla \gamma^2\, .
$$

We divide \eqref{eq.isentropic1} by $\gamma^2 \,w$ and, applying the above manipulations, obtain:
\begin{equation}
\frac {c^2}{w} \pmb\nabla w + \pmb v \times \Big[\pmb v \times \frac{\pmb \nabla w}{w}\Big] + \frac{c^2\, \pmb \nabla \gamma}{\gamma} - \frac{1}{\gamma^2} [\gamma\, \pmb v \times [\pmb \nabla \times \gamma \pmb v]] = - \pmb \nabla \Phi 
 \label{eq.isentropic2}
\end{equation}
Now let us scalarly multiply \eqref{eq.isentropic2} by $\pmb v$. 
Vectors $\Big [\pmb v \times \Big[\pmb v \times \frac {\pmb \nabla w}{w}\Big]\Big] $ and $[\gamma \pmb v \times [\pmb \nabla \gamma \pmb v]]$ are orthogonal to the velocity vector  $\pmb v$. Thus, their projections to the direction of the motion is zero and the scalar product of \eqref{eq.isentropic2} by $\pmb v$ yields:
\begin{equation}
\pmb v \cdot (\pmb \nabla \ln w + \pmb \nabla \ln \gamma + \frac{1}{c^2}\pmb\nabla \Phi) =0
\label{eq.bernoulli1}
\end{equation}
Taking into account the form for the gravitational potential $\Phi$  \eqref{eq.Phi_log}, we rewrite the last expression as 
$$
\pmb v \cdot \pmb\nabla \gamma\, w\, \Big(1-\frac{\Rg}{r}\Big)^{1/2} =0
$$

We thus obtain the following result. Along the flow lines, the following value is conserved:
$$
m_o\, c^2 \, w \gamma \Big(1-\frac{\Rg}{r}\Big)^{1/2} = m_o\, c^2 \, w \, \frac{\Big(1-\frac{\Rg}{r}\Big)^{1/2}}{\Big(1-\frac{v^2}{c^2}\Big)^{1/2}} = const
$$
This is a relativistic Bernoulli equation, written for the case of the Schwarzschild metric.

An elegant derivation of the relativistic { Bernoulli} equation, performed taking into account  the properties of the Killing vector field, can be found in \citet{Gourgoulhon2006,Gourgoulhon2007}.

\section{Summary}

The black hole gravitation causes the curvature of  space around it. A logarithmic potential can be introduced to describe the motion of particles in such gravitational landscape. { In contrast with  pseudo-Newtonian potentials, which can give only approximate results, the logarithmic potential provides the exact laws of motion. 
 For this,} we consider the logarithmic potential within a different approach, which represents the 3+1 { decomposition} of the Schwarzschild space-time near a black hole.  The advantage of such an approach for  GR problems is that it allows using the physical concepts analogous  to those in the classical physics. 
 
 In particular, the energy of a particle can be derived from the equation of motion using the logarithmic potential. We show that the  derived velocity of a particle, physically measured by a local observer, is correct in the sense  that it is  identical to that in GR. { The relativistic Bernoulli equation 
 for a fluid in the Schwarzschild metric is obtained.}

 %олагаю, что нужно подчеркнуть, что все зависит от задаваемой точности расчета задачи.
{ 
The choice of a potential and a method to deal with it depends on a desired accuracy of a problem. For considerations, which are not very precise, one can use the classical Newtonian mechanics and the Paczynski-Wiita's potential. It is not advised to use the logarithmic potential in the framework of the classical mechanics, since it gives less accurate results comparing to those obtained with the 
Paczynski-Wiita's potential (see discussion at the end of Sect.~\ref{s.velocities}). }
%(For example, for the logarithmic potential in the classical approach, the calculated value of the radius of the last stable orbit is $2\, \Rg$, meanwhile the exact value is $3\, \Rg$. Notice that the radius of the last stable orbit is $3\, \Rg$  for the Paczynski-Wiita's potential.) }

{ One can also use  the potential approach in the framework of classical mechanics to approximate the motion of a particle in the Kerr metric, by using a more sophisticated formula for a potential~\citep[see, for example,][where index $\beta$ is introduced]{kfm_book2008,artemova+1996}. However, the exact consideration of a particle motion in the Kerr metric implies the existence of a gravitomagnetic force, which is analogues to the Lorentz force in the electromagnetic theory and which is not conservative, that is, it cannot be determined by a potential~\citep[see, for example,][equations 3-18 and 3-19abc]{thorne_et1986}.
The exact force in the Kerr metric can be written out in the context of problem 1 of paragraph 88~in \citet{landau-lifshitz-e-2-1975} (see equation 3 there). This task could be a subject of another study.
}
 
\section*{Acknowledgments}
{ The authors are grateful to the anonymous referees for the  comments which helped substantially in our work on the manuscript.} The work is supported by the Russian Science Foundation grant 14-12-00146.

%\bibliographystyle{mnras}
%\bibliography{lipunova}

\end{document}